\documentclass[12pt]{article}
\usepackage[USletter]{vmargin}
\usepackage[T1]{fontenc}
\usepackage[]{times}
\usepackage{psfig,fancybox,fancyhdr,epsf}
\usepackage[]{fleqn}
\usepackage[dvips]{graphics}
\usepackage[english]{babel}
\begin{document}

\title{Non-equilibrium critical behavior : An extended irreversible thermodynamics approach}
\author{Enrique Hernández-Lemus$^{1}$\thanks{Corresponding author: hdz\_lemus@prodigy.net.mx}$\;$;  Leopoldo S. García-Colín $^{2, 3}$\\ {\small 1 Departamento de Física y Química Teórica, Facultad de Química, UNAM.} \\{\small Circuito Escolar s/n, Ciudad Universitaria, Coyoacán,  04510, México, D.F., México}\\ {\small 2 Departamento de Física, Universidad Autónoma Metropolitana-Iztapalapa} \\{\small Av. Purísima y Av. Michoacán s/n, 09340, México, D.F., México}.\\{\small 3 El Colegio Nacional, Luis González Obregón 23, México 06020, D.F., México}}
\date{}
\maketitle

\chead{Submitted to {\sl Journal of Non-Equilibrium Thermodynamics}}
\begin{abstract}
Critical phenomena in non-equilibrium systems have been studied by means of
a wide variety of theoretical and experimental approaches. Mode-coupling, renormalization group, complex Lie algebras and diagrammatic techniques are some of the usual theoretical tools. Experimental studies include light and inelastic neutron scattering, X-ray photon correlation spectroscopy, microwave interferometry and several other techniques. Nevertheless no conclusive treatment has been developed from the basic principles of a thermodynamic theory of irreversible processes.
We have developed a formalism in which we obtain
correlation functions as {\sl field averages} of the associated
functions. By applying such formalism we attempt to
find out if the resulting correlation functions will inherit the
mathematical properties (integrability, generalized homogeneity,
scaling laws) of its {\sl parent potentials}, and we will also use these correlation
functions to study the behavior of
macroscopic systems {\it far from equilibrium}, specially in the
neighborhood of critical points or dynamic phase transitions. As a working example we will consider the mono-critical behavior of a non-equilibrium binary fluid mixture close to its consolute point.
\end{abstract}


\section{Introduction}

The physical description of systems close to critical points has been attempted from  several standpoints ranging from the pure description of the critical opalescence of liquids by Andrews in the \textit{XIXth} century to modern approaches based on the application of Quantum Field theoretical tools such as various kinds of renormalization procedures \cite{binney}, non-affine algebraic structures, etc. Thermodynamically based descriptions of such systems are also available \cite{stanley, gw, widom65}. From the days of van der Waals, passing through Landau and Ginzburg, thermodynamic studies of critical properties have been related to geometrical properties of the thermodynamic state space. Different geometrical features were invoked: \textsl{accidental} geometrical relationships to describe coexistence or multi-criticality \cite{gw}; a Riemannian geometric conjecture relating the thermodynamic curvature with the correlation length and the free energy \cite{rupi}, etc.\\

Since critical phenomena involves strong dynamic behavior of fluctuations which induces abnormal transport, a non-equilibrium study will be adequate. Most dynamic studies of critical systems have been carried out based on the development of a description of internal modes known as mode-coupling theory. The pioneering works of Fixman \cite{fixman} and Kawasaki \cite{kawacrit} as well as others \cite{hohehal,halhohe,gunton, garcolmode} developed into a robust theory compatible with the experimental evidence from scattering and sound attenuation studies \cite{bursengers,bursengers2,changseng,bursenesf}. Unfortunately the mode-coupling framework commonly invokes the use of the so-called generalized hydrodynamics \cite{yipboon} based upon a linear (classical) version of the thermodynamics of irreversible processes. Linear irreversible thermodynamics assumes that the media under consideration could be taken as an equilibrium thermodynamic system pointwise, that is, that the quantities used to characterize an equilibrium state exist as continuous space-time functions. It is important to stress that not all thermodynamic systems could be viewed as these continua since in most cases there are domains or regions which either being composed of aggregates (like colloids and macromolecules on a suspension) or because of an uneven distribution of some property (anisotropic stress distribution, concentration domains, etc.) violate this pointwise equilibrium hypothesis which has been called the local equilibrium hypothesis \cite{lgcurib}.\\

 As it is known linear irreversible thermodynamics involves an assumption in the form of the already mentioned local equilibrium hypothesis stating that the description of systems close to equilibrium depends just on a local field evaluation of equilibrium quantities. Notwithstanding this condition mode-coupling theory often includes non-local effects due to the presence of non-equilibrium phenomena in the form of viscous and diffusive modes \cite{changseng}. These effects induce non-locality in space or time because of the presence of regions or domains. For example, in a system with concentration domains, mass diffusion occurs and so the concentration \textit{in one point} is changing in time, thus this point is not on an equilibrium condition. If the concentration gradients are small, as is the case when the equilibrium state is almost reached, one can find regions in which the concentration field is almost homogeneous and, in this case the local equilibrium will be a good approximation. Of course, this is not the case if the system presents very large values of fluctuations, since the relaxation of these fluctuations will induce domains that will produce transport processes, violating the local equilibrium assumption.\\

In order to deal with cases in which the local equilibrium hypothesis fails, several theories usually englobed under the name \textit{extended irreversible thermodynamics} have been developed. The common issue in all these extended irreversible thermodynamic theories is the inclusion of the non-local effects in the form of additional thermodynamic forces or, in some cases the addition of higher order terms in the gradients of the non-conserved properties. For example, if there is an uneven concentration distribution, this fact will induce mass transport. Mass transport as a non-equilibrium process will affect the whole thermodynamic description of the system, so that, for example its thermal diffusivity, its electrical conductivity and its heat capacity will change \cite{joucasleb}.\\

Due to the afore mentioned reasons in this work we attempt to introduce an extended irreversible thermodynamical description of the behavior of critical systems based on properties of the extended thermodynamic space (integrability, uniform stability, generalized homogeneity, etc.) that takes into account these large fluctuation non-local effects and in this way analyze the relation between criticality and dissipative phenomena.

\section{Thermodynamic formalism}

Following the tenets of Extended Irreversible Thermodynamics (EIT)  we shall start our discussion by assuming that a generalized entropy-like function $\Psi$ exists, which may be written in the form \cite{jourodgar}:

\begin{equation} {\frac{d \Psi^{\dagger}}{dt}} = T^{-1}[{\frac{dU}{dt}} +
 p{\frac{dv}{dt}} - \sum_i \mu_i^{\dagger}{\frac{dC_i}{dt}} -
 \sum_j {\cal X}_j:{\frac{d\Phi_j}{dt}}]\end{equation}

 or as a differential form

\begin{equation} \label{entrocal} d_t \Psi^{\dagger} = T^{-1}[d_tU + p d_tv -
\sum_i \mu_i^{\dagger}d_tC_i - \sum_j {\cal X}_j:d_t\Phi_j]\end{equation}

Equation \ref{entrocal} resembles the proposal made by Chen and Eu \cite{eumat} that

\begin{equation}\label{egibbs}T d_t \Psi = d_t U +  p d_tV - \sum_i {\mu_i^{\dagger} d_tC_i} + \sum_j{{\cal X}_j \odot d_t\Phi_j}\end{equation}
We see that equations ( \ref{entrocal} and \ref{egibbs}) are nothing but the formal extension of the celebrated Gibbs equation of equilibrium thermodynamics for the case of a multi-component non-equilibrium system. The quantities appearing therein are the standard ones,i.e. $T$ is the local temperature, $p$ and $V$ the pressure and volume, etc. $X_{j}$ and $\Phi_{j}$ are extended thermodynamical fluxes and forces. These extended forces and fluxes are the new elements of EIT, the ones that take into account the aforementioned non-local effects. In the case of a binary fluid mixture close to its critical consolute point we will take our set of relevant variables to consist in the temperature $T(\vec r, t)$ and concentration of one of the species $C_2(\vec r, t)$ as the slow varying (classical) parameters set $\cal{S}$ and the mass flux of the same species $\vec J_2 (\vec r, t)$ as a fast variable on the extended set $\cal{F}$ so that ${\cal{G}} = \cal{S} \, \bigcup \, \cal{F}$. These latter variable will take into account the presence of inhomogeneous regions (concentration domains in this case) and so will correct the predictions based on the local equilibrium hypothesis. The non-equilibrium Gibbs free energy for a binary fluid mixture under an isotropic mechanical environment ($dP = 0$) then reads:

\begin{equation} \label{negibb} d_tG = -\Psi d_tT + \mu^\dagger d_tC_2 - \vec X^\dagger \cdot d_t \vec J_2 \end{equation}

Here $\mu^\dagger = \mu_2 -\mu_1 $ is the so-called \textit{relative chemical potential}, $\vec J_2$ is the mass flux of species $2$ and  $\vec X^\dagger = \vec X_2 - \vec X_1$ is the thermodynamic driving force conjugated to this mass flux.\\

The non-equilibrium contribution to the generalized entropy is given (from a Maxwell-like relation) \cite{eumat} by:

\begin{equation} \label{maxy} \Psi_{ne} = \int_{0} ^{\vec J_2} ({{\partial \vec X ^\dagger}\over{dT}})\cdot d \vec J_2
\end{equation}

By substituting equation (\ref{maxy}) into equation (\ref{negibb}) and using a nonequilibrium Gibbs-Duhem relation \cite{bhalekar}a dynamic contribution to the concentration of species 1 is derived namely,

\begin{equation} \label{ce1} C_1^{*} =
[\int_{0} ^{T}{{C_p(T)}\over{T}}dT]({{\partial T}\over{\partial \mu ^\dagger}}) +
 [\int_{0} ^{\vec J_2} ({{\partial \vec X ^\dagger}\over{dT}})\cdot d \vec J_2]
 ({{\partial T}\over{\partial \mu ^\dagger}}) - \vec X ^\dagger \cdot
 ({{\partial \vec J_2}\over{\partial \mu ^\dagger}})\end{equation}

If we now assume that Antoine's law $\mu ^\dagger (\vec r,t) = \mu ^{\dagger ^0} + H(\vec r) T(\vec r,t)$ holds locally, $H(\vec r)$ is a time independent parameter playing the role of the constant R  (ideal gas constant) of the equilibrium Antoine's law; then we can derive a closed expression for $C_1^{*}$ for a given set of constitutive equations for the fast variables (in this case the mass flux and its driving force). We propose linear constitutive equations with exponential memory kernel for the mass flux $\vec J_2$ and its conjugated force $\vec X ^\dagger$ for the following reasons:
\begin{enumerate}
\item The associated transport equations are hyperbolic (of the Maxwell-Cattaneo-Vernotte type) \cite{cattaneo, vernotte} so causality is taken into account.
\item These hyperbolic transport equations are compatible with the postulates of EIT  \cite{joucasleb,leboncas, muschik, muschikasp, muschikrec}
\end{enumerate}

The constitutive equations are therefore chosen to be,
\begin{equation} \label{jota1} \vec J_2 (\vec r,t) =  \int_{-\infty} ^{t} \lambda_1 \, \vec u \, e ^{\frac {(t'-t)}{\tau_1}}
\mu ^\dagger (\vec r,t') dt'\end{equation}

\begin{equation}\label{equis1} \vec X ^\dagger (\vec r,t) = \int_{-\infty} ^{t} \lambda_2 \, e ^{{(t''-t)}\over{\tau_2}}
\vec J_2 (\vec r,t'') dt'' \end{equation}
                                                                                                The $\lambda_i$'s are time-independent, but possibly anisotropic amplitudes, $\vec u$ is a unit vector in the direction of mass flow (the nature of $\vec u$ will not affect the rest of our description, since we will be dealing with the magnitude of the mass flux $|\vec J_2|$) and $\tau_i$'s are the associated relaxation times considered path-independent scalars. Since we have a linear relation between thermodynamic fluxes and forces some features of the Onsager-Casimir formalism  will still hold.\\

By using linear constitutive equations we are limiting our treatment to comparatively small spatial variations, otherwise we will have to make a gradient expansion, namely $J \sim \nabla \mu$, or more generally $J = \sum_{n} \epsilon_n \nabla^{n} \mu$, where $\epsilon_n$ are the expansion coefficients. We believe that introducing the gradient expansion of the constitutive equations will not supply a big deal of new physics since the first contribution will be the most important from the standpoint of thermodynamics. We are instead considering at a deeper level the dynamic effect given by the exponential memory kernel. We have already talked about two fundamental approaches in EIT, one is expanding the gradients of the non-locally conserved densities; while the other consists in enlarging the set of relevant parameters by adding nonequilibrium fluxes and forces \cite{joucasleb}. In this work we choose to take the latter approach by considering it more appropriate in dealing with dynamic critical phenomena for the reasons just mentioned. We did just so in Eqns. \ref{maxy}, \ref{ce1} and \ref{equis1}.\\

As we can see, equations (\ref{jota1}), (\ref{equis1}) together with Antoine's law define a dynamical coupling between the thermal field and the concentration field. We will take the case in which the temperature field decays exponentially (a so-called multiplicative thermostat) since this is a common situation in experimental studies near the critical point \cite{expe}. Therefore,

\begin{equation} \label{expoter} T = T_{eq} + T(\vec r) \, e^{-nt} \end{equation}

where $T$ is the non-equilibrium local temperature, $T_{eq}$ is the temperature associated with the \textit{target} equilibrium state, namely, the closest equilibrium state to the critical point and the one towards which the system relaxes in time, $T(\vec r)$ is an effective amplitude for the thermal decay rate and $n$ is a dynamic critical exponent related to thermal relaxation whose explicit nature will be evidenced later by considering its relation with the distribution of decay rates of concentration fluctuations. Substitution of equation \ref{expoter} into Antoine's law, makes it possible to calculate the first derivative in equation \ref{ce1}, namely $({{\partial T}\over{\partial \mu ^\dagger}})$, and by substituting Antoine's law in equation \ref{jota1} and then $J_2$ into equation \ref{equis1} we are able to write down $({{\partial \vec X ^\dagger}\over{dT}})$. In a similar fashion one can obtain
$({{\partial \vec J_2}\over{\partial \mu ^\dagger}})$, etc. After substituting the derivatives and solving the integral over $\vec J_2$, the local expression for the concentration (equation \ref{ce1}) reads:

\begin{eqnarray}\label{concentra} C_1^{*} (\vec r,t)& = & [\int_{o} ^{T}{{C_p(T)}\over{T}} dT]({{1}\over{H(\vec r)}}) +
\nonumber \\ && [{{\lambda_1 \lambda_{2}  \tau_1 H(\vec r) T(\vec r)}\over{n(1-n \tau_1)
^2}}][\lambda_1 \tau_1 ^2 -  \tau_2 ({{\lambda_1 \tau_1}\over{(1-n \tau_1)}} \nonumber \\ && +
{{ \lambda_1 (1-n \tau_1) -1}\over{(1-n \tau_2)}}) ]\, e ^{-nt}\end{eqnarray}

In order for the integrals to be convergent and physically sound, some relations arise between the relaxation times $\tau_1$ and $\tau_2$ and the nonequilibrium critical exponent $n$, namely $n \tau_1 < 1$ and $n \tau_2 < 1$. It is through this kind of relations that an irreversible thermodynamics study results enlightening, since the connection between criticality and relaxation times is evidenced.\\

\section{Generalized homogeneous functions and scaling hypotheses}

\subsection{Static scaling hypothesis}

In the neighborhood of a critical point one of the relevant fields (the locally conserved  densities and the order parameter $\psi$) exhibits a stronger divergence than the others. This highly divergent field called the {\it order parameter field} has an associated field  $h_{\psi}$ that defines the {\it generalized susceptibility}. The so-called scaling hypothesis is the assumption stating that the dominant singularity in the free energy
$F(h_{\psi}, T)$ may be rephrased in the following form (at least asymptotically):

\begin{equation} \label{widfish} \frac{F(h_{\psi}, T)_{sing}}{V k_BT_c}= f^{*}(h_{\psi}, T)= A_0 |\Delta T|^{2-\alpha}
f\left(\frac{K_0 \Delta T}{|h_{\psi}|^{\frac{1}{\beta\delta}}}\right) \end{equation}
The scaling theory of static critical phenomena \cite{stanley} states that the generalized susceptibility associated with the order parameter follows a power-law:
\begin{equation} \Lambda_\psi \sim \zeta^{2-\eta} \end{equation}

$\Lambda_\psi$ is the susceptibility associated with the order parameter, $\zeta$ is the correlation length and for the binary fluid mixture the associated universality class is that of the isotropic 3-D Ising system ($d=3$ dimensions, $n=1$ implying a scalar order parameter (concentration difference)). For this universality class it has been found that $\eta = 0.04$ and that $ \zeta \sim (T_{c} -T)^{-\nu}$ with $\nu \simeq 0.62$ \cite{stanley}.\\

\subsection{Dynamic scaling hypothesis}

The dynamic scaling hypothesis (DSH) states that in the neighborhood of the critical point a scaling factor $x$ given by  $x= q  \zeta$ remains finite notwithstanding the divergent correlation length $\zeta$, due to the fact that in the thermodynamic limit the wave number vanishes ($ q \to 0$). On the other hand, according with the theory of dynamic critical phenomena
\cite{hohehal, gunton} the static structure factor $\chi_q$ follows a power law:

\begin{equation}\chi_q=\Gamma \epsilon^{-\gamma}g(q \zeta)\end{equation}

with $\epsilon$ =$ (T-T_c) \over T_c$,   $g(x)$ an unspecified monotonic function and $\zeta$ the correlation length, that goes as $\zeta = \zeta_0 \epsilon^{-\nu}$.\\

This concentration fluctuations decay rate has been related to {\sl viscous modes} by Sengers et al \cite{bursengers, bursengers2,bursenesf} giving the dynamic scaling exponent (in Fourier space) for the non-regular part of the diffusion coefficient from linearized mode-coupling theory as:
\begin{equation} \label{dshd} D = D_0 q^y \end{equation}

where $y = 1+z_\eta \simeq 1.0678$, $z_\eta$ being the critical exponent related to the asymptotic non-regular shear viscosity.\\

The diffusion coefficient exhibits anisotropy $D=D(q)$ but according with the DSH a scaling factor $x$ exists such that $q$ = $x \over \zeta$ so that $D = D_0 \, q^y$  gives $D=D_0$$ x^y \over \zeta^y$, $x$ remaining finite so $x^y$ may be incorporated in the amplitude $D_0$ hence, $D=D_1 \zeta^{-y}$ which clearly vanishes at the critical point where $\zeta$ diverges. In this case one has:

\begin{equation} \label{difu1} D = D_1(\zeta_0 \epsilon^{-\nu})^{-y} =
D_2 \epsilon^{\nu y}  = D=D_2 \epsilon^{\beta'}\end{equation}

                                                                                            Equation (\ref{difu1}) gives us the asymptotic dependence of the diffusion coefficient with temperature in the vicinity of the critical point via the exponent $\beta'$. By means of scaling relations for the non-equilibrium thermodynamic potentials (based on their property of generalized homogeneity) the critical exponent for this transport coefficient will be calculated. Once this value for $\beta'$ is known, we will be able to calculate $\nu$, since $\nu = \frac{beta '}{y}$. The value ontained for the exponent $\nu$ could be compared with the DSH-mode coupling result in order to establish compatibility criteria.  \\

We will begin by analyzing the scaling hypothesis in the representation of the symmetric thermodynamic potential ($f^{*}$) for a non-equilibrium binary mixture described by the time-dependent fields temperature (T), concentration of one species ($C_i$) and mass flux of this species ($\vec J_i$):

\begin{equation}\label{simi} f ^*=f^ *(T,C_i,\vec J_i) \end{equation}

Equation (\ref{simi}) is a non-equilibrium extension of equation (\ref{widfish}), the order parameter is given by the concentration field $C_i$, and the non-equilibrium contribution is given by the mass flux $\vec J_i$.  We will assume here that (generalized) mono-scaling applies :

\begin{equation} f^{*} (\lambda^{a_E} \epsilon, \lambda^ {a_c} C^{+}, \lambda ^{a_J} \vec J_i)=$ $\lambda f^{*} (\epsilon , C^+, \vec J_i)\end {equation}

where $\epsilon$ = ${{T - T_c} \over {T_c}}$; $C^+$ = ${{C_1^I- C_1^{II}} \over {C_c}}$, $C_c$ the critical consolute composition and $\lambda$ an arbitrary parameter. I and II refer to the corresponding phase. Partial derivatives of $f^*$ respect to its natural variables are conjugated.

\begin{eqnarray} ({\frac {\partial f^*}{\partial C^+}}) = \mu^+(\epsilon , C^+, \vec J_i); &&  ( {\frac{\partial f^*}{\partial \vec J_i}}) = - \vec X_i (\epsilon , C^+, \vec J_i) \nonumber \\  \mu^+ =  {\frac{ \mu^\dagger- \mu^\dagger_c}{\mu^\dagger_c}}; && \mu^\dagger = \mu_1 - \mu_2.\end{eqnarray}

Since $f^*$ is considered to be a generalized homogeneous function, it follows that:

\begin{equation} \lambda^{a_J-1} \vec X_i(\lambda^{a_E} \epsilon, \lambda^ {a_c} C^+, \lambda ^{a_J} \vec J_i) = \vec X_i (\epsilon , c^+, J) \end{equation}

\begin{equation} \lambda^{a_c-1} \mu^+ (\lambda^{a_E} \epsilon, \lambda^ {a_c} C^+, \lambda ^{a_J} \vec J_i) =  \mu^+ (\epsilon , C^+, \vec J_i) \end{equation}

which implies the following scaling relations:

\begin{eqnarray} \mu ^+ = \mu_0 (-\epsilon)^\beta; \; X= X_0 (-\epsilon)^{\beta '} \nonumber \\ \mu ^+ = \mu_0 (C^+)^\delta ; \;   X=X_1 (C^+ )^{\delta '} \nonumber \\ \mu ^+ = \mu_0 (J)^r ; \; X=X_2 (J)^{r '} \end{eqnarray}

From now on we will obviate the subscripts in $\vec J_i$ and $\vec X_i$ and we will work with their scalar magnitudes since we are interested in the thermal effect of the mass flow given by these scalar magnitudes. Simple relations between the exponents may be derived as it is usual \cite{lgctermo}; namely,

\begin{eqnarray} \label{expo1} \beta =  {{1-a_c} \over {a_E}}; \;  \beta'  =  {{1-a_J} \over {a_E}}; \; \delta  =  {\frac{1-a_c}{a_c}}; \; \delta'  =  {\frac {1-a_J}{a_c}}; \; r  =  {{1-a_c} \over {a_J}}; \; r'  =  {{1-a_J} \over {a_J}} \end{eqnarray}

These equations may be casted to give relations between exponents related to either non-equilibrium or equilibrium quantities (Remember that partial derivatives as $\mu ^+$ are calculated at constant values of J, that is steady states that can made be coincident with reversible states ($\Psi \to S$)) so we get:

\begin{eqnarray} \label{relac} a_c = {{1}\over{1+ \delta}} = {{r'}\over{\delta'(1+r')}}; \; a_J = {{1}\over{1 +r'}} = {{\delta}\over{r(1+\delta)}}; \; a_E = {{\delta}\over{\beta(1+ \delta)}} = {{r'}\over{\beta'(1+r')}}\nonumber \end{eqnarray}

As we have seen {\sl equilibrium} quantities  are now being supplemented with a non-equilibrium exponent in order to determine the others, due to the fact that only 3 of these exponents are independent, two of them from equilibrium and one for the nonequilibrium processes. In the case that concerns us, we want to get relations leading to the determination of exponent $\beta'$ that, as is clear from Eq. \ref{expo1} is given in terms of $a_J$ and $a_E$.  Eqns. \ref{relac} show that $a_J$ and $a_E$ are related to either the two equilibrium exponents $\beta$ and $\delta$ in the case of $a_E$ or to the non-equilibrium exponent $r'$ in the case of $a_J$.\\

Under the particular approximation made in this work (namely, linear but causal processes) , the constitutive equation \ref{equis1} implies the value $r'=1$. The validity of this assumption in the context of dynamic critical phenomena is left for an \textit{a posteriori} analysis based on the consequences in the final results. Nevertheless, given the dynamic character of critical phenomena, we believe that the temporal component (taken into account in the memory equation Eq. \ref{expo1} for the nonequilibrium contribution) will be of greater importance than the spatial inhomogeneities. For these reasons we assume the dynamic critical exponent to have the value $r'=1$ and relate it with the two additional equilibrium exponents in order to make the following calculations.\\

The diffusion coefficient behaves as a power-law with the reduced temperature $\epsilon$ with an exponent $\beta '$ as:\\

\begin{equation} \label{difubeta} D \nabla c = J \,;  \;   J \sim \epsilon^{\beta '} \;  \to D \sim \epsilon^{\beta '} \end{equation}

Equation (\ref{difubeta}) has been derived from the scaling procedure of the extended thermodynamic potentials. If we recall equations (\ref{expo1}) it has been possible to see that only three of the exponents are independent. Having noticed this and being aware of the values of $\beta$ and $\delta$ from equilibrium thermodynamics, as well as the value of $r'=1$ from our linear constitutive relation [equation (\ref{equis1})] it is possible to find out the value of $\beta '$ and from it the value of the exponent $\nu$. Since we are interested in a complete thermodynamical characterization of critical dynamic phase transitions we will also look at the behavior of the associated correlation function.

\section{Non-equilibrium correlation functions}

It has been proposed that macroscopic correlation functions may
be obtained from a non-equilibrium thermodynamical description, by using the fact that extended thermodynamical potentials belongs to a certain class of statistical systems whose probability distribution functions obey a certain relaxation
times hierarchy on its evolution towards the most
probable state.\\

We will use the equivalence between 2-time correlation functions and time evolution operators acting on a dynamical variable \cite{zwanzigop} in order to calculate these correlation functions. The {\sl dynamic variables} are the set of extended thermodynamical potentials. {\sl Time evolution operators} are obtained as inner products of the aforementioned thermodynamical potentials. If the thermodynamic average is performed under isotropic canonical conditions \cite{zubasm, zubarusm, hurley} the resulting correlation function is given by: \\

\begin{equation} \label{nneitvh}
\langle C_1(\vec r,t) , C_1(\vec r,t') \rangle = \int_{\Phi} \Xi(\vec r) \; e ^{-nt} \; e ^{-n' t'} \; W(t- t')dt' \end{equation}

At this point it becomes necessary to clarify the meaning of the dynamic exponent $n$. Exponent $n$ is, in principle, a time-dependent quantity characterizing the decay rate of the thermal field at time $t$ and $n'$ its value at time $t'$. In the approximation taken in this work this decay rate is closely related to the decay rate of concentration fluctuations $\Gamma$. For similar systems a distribution of decay rates $\Gamma (t)$ has been proposed \cite{bursengers,bursengers2} giving in some cases, a double exponential decay behavior for the correlation function.\\

Nevertheless, in the case of a simple mixture (non-colloid, non-micellar, etc.) a simple exponential fits well the experimental data. This is accomplished if we consider the dynamic exponent $n$ to have a weak time dependence, so that $n = n'$ at least along the integration path $\Phi$. In doing this we are excluding crossover phenomena, as well as some features of complex fluid behavior of our analysis. It is worth mentioning that recently, some complex fluids have been shown to follow this exponential decay of the correlation function. We will mention the case of an ionic binary (ethylammonium nitrate-n-octanol) mixture \cite{mirza}, and also binary mixtures of small molecular weight fluids (like hexane-nitrobenzene) \cite{dufre}. In contrast, a  protein in aqueous solution has been observed to significantly deviate form exponential behavior \cite{fine}. In the single exponential approximation one has:

\begin{equation} \label{eitvh}
\langle C_1(\vec r,t) , C_1(\vec r,t') \rangle = \int_{\Phi} \Xi(\vec r) \; e ^{-n (t+t')} \; W(t - t')dt' \end{equation}

here $W(t - t')$ is a weighting function and the following abbreviations have been made,

\begin{equation}
\Xi (r) = AB + B ^2
\end{equation}
\begin{equation} A = \int_{o} ^{T}{\frac{C_p(T)}{T}} dT \, {\frac{1}{H(\vec r)}} \end{equation}

\begin{equation}
B=[{{\lambda_1 \lambda_{2}  \tau_1 H(\vec r) T(\vec r)}\over{n(1-n \tau_1) ^2}}]
[\lambda_1 \tau_1 ^2 - \tau_2 ({{\lambda_1 \tau_1}\over{(1-n \tau_1)}}+{{ \lambda_1 (1-n \tau_1) -1}\over{(1-n \tau_2)}}) ]
\end{equation}

Equation (\ref{eitvh}) is the non-regular part of the composition field time correlation function.

One of the advantages of this formalism with respect to others, such as mode-coupling theory  is that we can test different kinds of relaxational couplings (delta-correlated, quasi-markovian, gaussian,  lorentzian, Ornstein-Uhlenbeck type,
non-markovian, simultaneous non-linear multi-modal coupling, etc.) just by changing the weighting functions $W(t - t')$ (that is the family of probability measures), within a unifying thermodynamic scheme.\\

For a broad family of weighting functions, equation (\ref{eitvh}) asymptotically converges to a limit given by \cite{ryzhik}:

\begin{equation} \label{convh}
\langle C_1(\vec r,t) , C_1(\vec r,t') \rangle =  \kappa \; \Xi(\vec r) \; e ^{-2n t}  \end{equation}

where $\kappa$ is a constant depending on the explicit weighting function . For an Ornstein-Uhlenbeck measure ($W(t-t') = e^{|t-t'|}$) convergence implies $\kappa = {\frac {1}{1-n}}$. In the case of Gaussian measure ($W(t-t') = e^{(t-t')^{2}}$) an asymptotic solution similar to equation (\ref{convh}) could not be given in terms of an exponential decay. In this case the solution consist of two contributions: an exponential decay plus an error function type mode. This difference eliminates the short-time plateau (known as critical slowing down of fluctuations) present in correlation functions obtained using colored noise measures as weighting functions.

\section{Results and discussion}

Once we have been able to calculate the correlation function, we are now in position to discuss some interesting physical results derived from the scaling analysis of this function and its comparison with experimental data.

\subsection{Diffusion coefficient critical exponent}

The concentration correlation function (as obtained by light scattering or
inelastic neutron scattering) is given by the following (van Hove) expression :

\begin{equation} \label{vanthove}\langle C_1(\vec r, t) \, C_1(\vec r, t') \rangle = \chi (\vec r) e^{-D q^2 t}  \end{equation}

It is a known fact that the dynamics of the fluctuations {\sl slow down} close to a critical point. If we look at the critical consolute point of a binary fluid mixture
a sound physical reason for this critical slowing down could be extracted from a simple analysis of equation (\ref{vanthove}) for the 2-time concentration correlation function. In this case the diffusivity D is to be identified with the mass diffusion coefficient for one of the species between the two phases. At the consolute point (where the two phases coalesce in one single phase) there is no driving force, for the diffusion mechanisms since spatial concentration gradients no longer exist.\\

Considering an exponential thermal decay (a multiplicative thermostat) from the critical
point [equation (\ref{expoter})] we prepare the system at the critical point at $ t = 0$  with  the critical temperature given by:

\begin{equation} T_c = T_{eq} + T(\vec r) \end{equation}

The correlation length diverges according with the power law:

\begin{equation} \zeta \sim ({\frac{T_c}{T_c - T}})^\nu \end{equation}

In our case:

\begin{equation} \label{correle} \zeta = \zeta_0 ({\frac{T_c}{T(\vec r)(1-e^{-nt}) }})^\nu \end{equation}

at t=0 (critical point)  $T=T_c$ y $\zeta \to \infty$.\\

In the long-time limit the system will reach an equilibrium state (the {\it target} state)
with the following properties, at $t \to \infty$,  $T=T_{eq}$, $\zeta = \zeta_0 \,({\frac{T_c}{T(\vec r)}})^\nu$.The physical meaning of this equilibrium state and its associated temperature and correlation length will become clearer later.\\

Let us recall equation \ref{eitvh}, since the critical concentration correlation function decays as $e ^{-2nt}$, for a variety of non-equilibrium measures $W(t')$ [as stated by equation (\ref{convh})] we get the following relations:

\begin{equation}\langle C_1(\vec r,t) , C_1(\vec r,t') \rangle \sim e ^{-2nt} \sim  e^{-D q^2 t} \end{equation}

It has been proved useful to define a decay rate of fluctuations $\Gamma = D q^2$ according to equation \ref{vanthove} as an indicator of the decay velocity of the concentration correlations. This definition clarifies the meaning of our dynamic critical exponent $n$ since $\Gamma = 2n$. Following the behavior of the diffusion coefficient given by Sengers et al \cite{bursengers2,changseng,bursenesf} namely equation (\ref{dshd}), we have:

\begin{equation} e ^{-2nt} =  e^{-D_0 q^{y+2} t};   \, \; n= {\frac {D_0 q^{y+2}}{2}} \end{equation}

\begin{equation} n= \left({\frac{D_0}{2}} \right) \left({\frac{x}{\zeta}}\right)^{y+2} \end{equation}

but the correlation length depends on the time-dependent temperature field according to equation (\ref{correle}) hence,

\begin{equation} n=  \left({\frac{D_0}{2}} \right) \left({\frac{x}{\zeta_{0}({\frac{T_c}{T(\vec r)(1-e^{-nt}) }})^\nu}}\right)^{y+2} \end{equation}

and so,

\begin{equation} \label{ene} n= {\frac{D_0 (x/\zeta_{0})^{y+2} T(\vec r)^{\nu ({y+2})} (1-e^{-nt})^{\nu ({y+2})}}{2 T_c ^{\nu ({y+2})}}} \end{equation}

Since the decay rate of concentration fluctuations $\Gamma = 2n$ equation \ref{ene} can be rephrased:

\begin{equation} \label{decay} \Gamma = {\frac{D_0 (x/\zeta_{0})^{y+2} T(\vec r)^{\nu ({y+2})} (1-e^{-nt})^{\nu ({y+2})}}{T_c ^{\nu ({y+2})}}} \end{equation}

The critical exponent $\nu$ may therefore be obtained from the thermodynamical description given by equation (\ref{ene}) as follows:

\begin{equation} \label{nunu}  \nu = \lim_{t \to \; 0} \; \;({\frac{1}{y+2}}) {\frac{ln \left({\frac{D_0 (x/\zeta_{0})^{y+2}}{2n}}\right)}{ln\left({\frac{T_c}{T(\vec r)(1-e^{-nt})}}\right)}} \end{equation}

Equation \ref{nunu} is the main result of this paper, by means of it and by multiplying by $y = 1+z_\eta$ a value for $\beta'$ is obtained to compare with the value of equation \ref{difu1}. The value for $\nu$ was obtained by numerical evaluation of the limit ($\lim_{t \to \; 0}$). We made this by evaluating the function for several values of $t$ and then by imposing an error bound of $0.001$ in the quantity calculated. We looked up for the accumulation point of the function near its zero-time limit, that is, we use a Cauchy convergence criteria in the limit. This leads us to adjust a cut-off time value of the order of $95 \; milliseconds$. Our closest approach to the limit given by this value of $t$. Given this facts, we should expect an error bounded in the third significant decimals of exponent $\nu$ since Cauchy criteria is linear on the error bound.\\

It is possible to rearrange the preceding expression (Eq. \ref{nunu}) to give:

\begin{equation} \label{omegaw}\left({\frac{2nT_c^{\nu (y+2)}}{D_0 T(\vec r)^{\nu (y+2)}}}\right) = (x/\zeta_{0})^{\nu (y+2)}(1-e^{-nt})^{\nu (y+2)}
\end{equation}

We now define $\Omega$, and call it the modified dynamic Kawasaki function for reasons to be seen, as follows:

\begin{equation} \label{kawao} \Omega (t) = \left({\frac{\Gamma \, T_c^{\nu (y+2)}}{D_0 T(\vec r)^{\nu (y+2)}}}\right)^{\frac{1}{(y+2)}}
\end{equation}

Substitution of equation \ref{omegaw} into equation \ref{kawao} yields finally that,

\begin{equation} \Omega (t)  = (x/\zeta_{0}) \,(1-e^{-nt})^{\nu} \end {equation}

We wrote $\Omega$ as a function of time via the decay rate $n$ and not just as a function of the scaling parameter $x$ in order to avoid the use of the  Ornstein-Zernike (OZ)  approximation. We notice that $\Omega$ is finite and regular at $t=0$ (Critical point) and in the  $t \to \infty$ limit (thermodynamic equilibrium).

\begin{equation} \Omega (t=0) = 0 ; \; \; \; \; \; \Omega (t \to \infty) = (x/\zeta_{0}) \end{equation}

This modified Kawasaki function (scaled by the exponent $\varphi = {\frac {1}{y+2}}$) possess a clearer physical interpretation than the one obtained from the asymptotic Ornstein-Zernike form of the radial distribution function $g(x)= {\frac{1}{1+x^2}}$ as treated by means of Mode Coupling Theory for the case of non-equilibrium states since the closure of the OZ equation is given by a local equilibrium distribution.\\

If we take experimental values for the amplitudes ($D_0$, $T(\vec r)$) and other characteristic parameters \cite{dufre, sales}, it is possible to express a convergent sequence leading to the determination of $\nu$ from equation (\ref{nunu}). In order to calculate $\nu$ we had to adjust a \textsl{cut-off time} of the order of $95 \; ms$ (or a cut-off frequency of  10.52 Hz). To within this approximation we obtain an average value of $\nu_{calc}=0.6303$ for the set of realizations for the limit in equation \ref{nunu}. Once this critical exponent is known we proceed to express the behavior of other physically relevant quantities.\\

In fig. 1 we plotted the predicted behavior of the mass diffusion coefficient $D$ in the proximity of the critical point. According with the DSH treatment in the already mentioned mode-coupling description \cite{bursengers} $D=D_2 \, \epsilon^{\nu \,y} =D_2 \, \epsilon^{\beta'} $, if we use our calculated value $\nu_{calc}=0.6303$ we get the solid line plotted in fig. 1. We also plotted experimental results for an ionic binary mixture as reported from recent scattering experiments \cite{sales}. As is shown good agreement between experiment and theoretical predictions is achieved even for this case (an ionic mixture). We must anyhow say a word of caution about the effect of using a reduced temperature coordinate since this fact stretches the curve and thus gives a visual impression of higher accuracy. What is surprising is that even for this relatively simple model (that leaves out many characteristic features) essential behavior is reproduced: a signature of a thermodynamic analysis.\\

In fig. 2 the wave number dependency of the diffusion coefficient is shown. From some time ago experimentalists have claimed a non-fickian behavior near the critical point against some theoretical assumption of linearity derived from generalized hydrodynamics. Fig. 2 shows that even though this deviation from Fick's law exists indeed the effect is really small. Our formalism predicts $D \sim q^{y}$ with $y=1.0678$. Since this was obtained using the mono-scaled DSH it is also possible that further corrections need to be taken into account in the future.  This fact will not affect significantly our consideration made in equations \ref{equis1} regarding $r' = 1$.\\

In fig. 3 we show that the decay rates of concentration fluctuations deviates slightly from the parabolic behavior given by linear response predictions. Fig. 3 shows a semi-flat region in the low wave number regime that possibly indicates a macroscopic effect of stability at long distance scales.\\

In Table 1 we compare our results with some of the most reliable data published in the recent literature of dynamic critical phenomena. The value of exponent $\nu$ is somewhat higher than the reported values. This fact may be explained because the experimental data we used for our calculations were measured on a region that, appart from critical phenomena, exhibits other complex dynamic processes like Rayleigh-Bénard convection due to temperature fluctuations. Our calculation of $\nu$ is based on the behavior of the decay rate of thermal fluctuations $\Gamma$ according to equations \ref{decay} and \ref{nunu}. A cross-over in the associated structure factor's dependence of the wave number $q$ is known to exist on passing from large scale to small scale, and this fact has been related with the presence of two modes in the Rayleigh spectrum, one thermal mode identified with thermal fluctuations of decay rate $D_Tq^{2}$ and one viscous mode with a decay rate $\lambda q^{2}$ \cite{oh}. Here $D_T$ is the thermal diffusivity and $\lambda$ is the kinematic viscosity. In our theoretical approach we did not take viscous effects into account except in the consideration for exponent $y = 1 + z_{\eta}$ in equation \ref{dshd} and so viscous dissipation is disregarded in our treatment of the dynamics of thermal fluctuations. Since the viscous mode is faster, the difference in the dynamics between this calculation and others should vanish eventually, although an effect in the critical exponent, even if small, is present. \\

An additional analysis of fig. 3 suggest that it is possible that, just as there is a dynamic effect of slowing down of critical fluctuations, there is also a spatial effect of smoothing of critical fluctuations. Further investigations are need in the future in this direction and, possibly by inclusion of other dissipative effects (viscous modes notably). Our physical intuition about the applicability of non-equilibrium thermodynamic description to critical behavior will be substantially increased.

\newpage

{\bf \large Figure captions}\\

{\bf Figure 1} Temperature dependence of the near critical mass diffusion coefficient. Solid line is this works prediction [equation (\ref{difubeta})]. Square dots and open circles are experimental values taken from reference \cite{sales}.\\

{\bf Figure 2} Wave number dependence of the near critical mass diffusion coefficient.
A slight deviation of Fick´s law could be noted. The curve represents a power-law behavior $D \sim q ^{1.0678}$.\\

{\bf Figure 3} Decay rate of fluctuations $\Gamma$ as a function of the wave number. This work predicts a slight deviation from the linear response parabolic behavior. It is noticeable the presence of a plateau in the low wave number region which possibly represents a macroscopic effect of spatial smoothing down of fluctuations.

\newpage

\begin{tabular}{|c|r|r|}
\hline
Critical exponent&Theoretical calculations &Renormalization \& simulations \\
\hline
$\nu$& \textbf{0.6303} $\pm$ \textbf{0.001} \textbf{This work} & -\\
\hline
\hline
$\nu$& 0.63012(16) HTS \cite{campo1}& 0.618 \cite{campo2}\\
\hline
$\nu$& 0.6302(2) HTS \cite{campo1}& 0.630 $\pm$ 0.00015 \cite{guida}\\
\hline
$\nu$& 0.63002(23) HTS \cite{campo2}& 0.632 $\pm$ 0.003 \cite{guida}\\
\hline
\end{tabular}


 Como vemos en la tabla \ref{lalala} esto es chido.
\newpage

\begin{figure}[h]
\centerline{\psfig{file=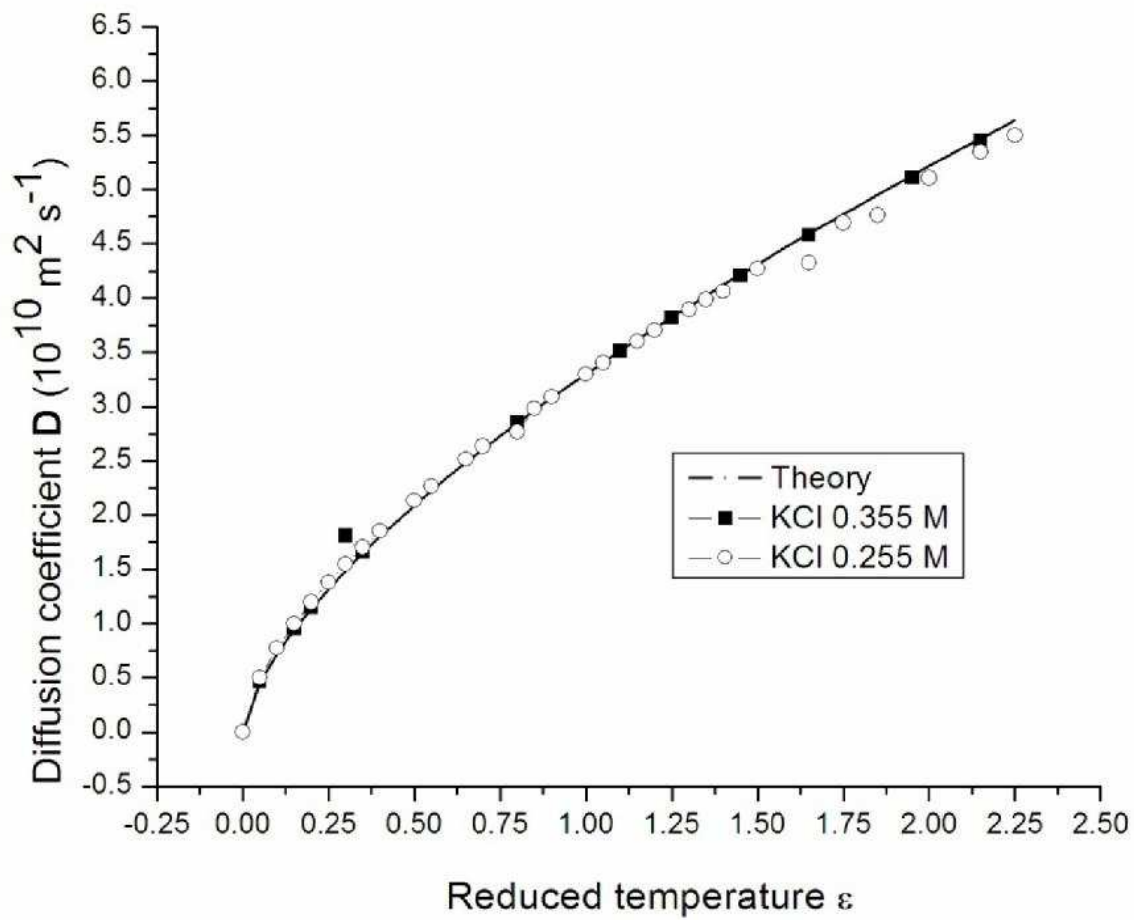,width=18cm,angle=0}}
\caption{Temperature dependence of the near critical mass diffusion coefficient. Solid line is this work's prediction [equation (\ref{difubeta})]. Square dots and open circles are experimental values taken from reference \cite{sales}}
\label{Figure 1}
\end{figure}

\begin{figure}[h]
\centerline{\psfig{file=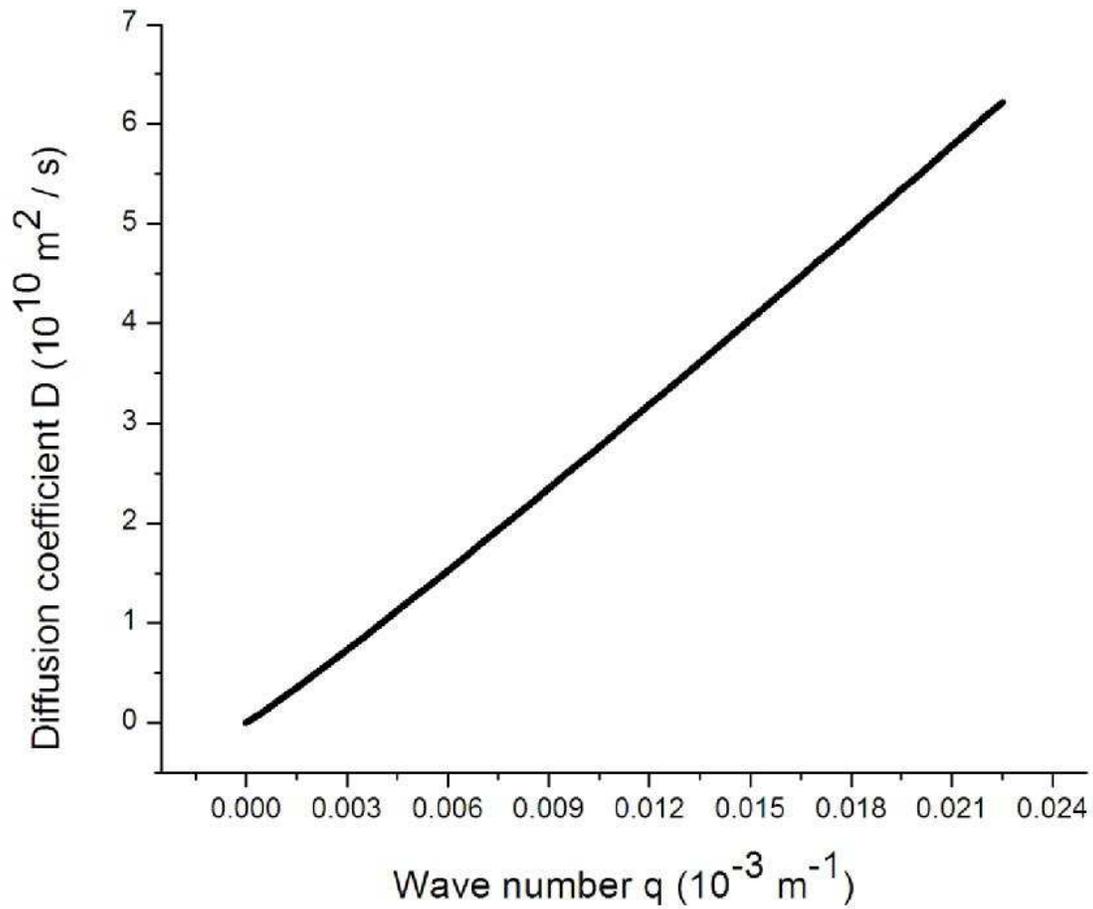,width=18cm,angle=0}}
\caption{Wave number dependence of the near critical mass diffusion coefficient. A slight deviation of Fick´s law could be noted. The curve represents a power-law behavior $D \sim q ^{1.0678}$.}
\label{Figure 2}
\end{figure}

\begin{figure}[h]
\centerline{\psfig{file=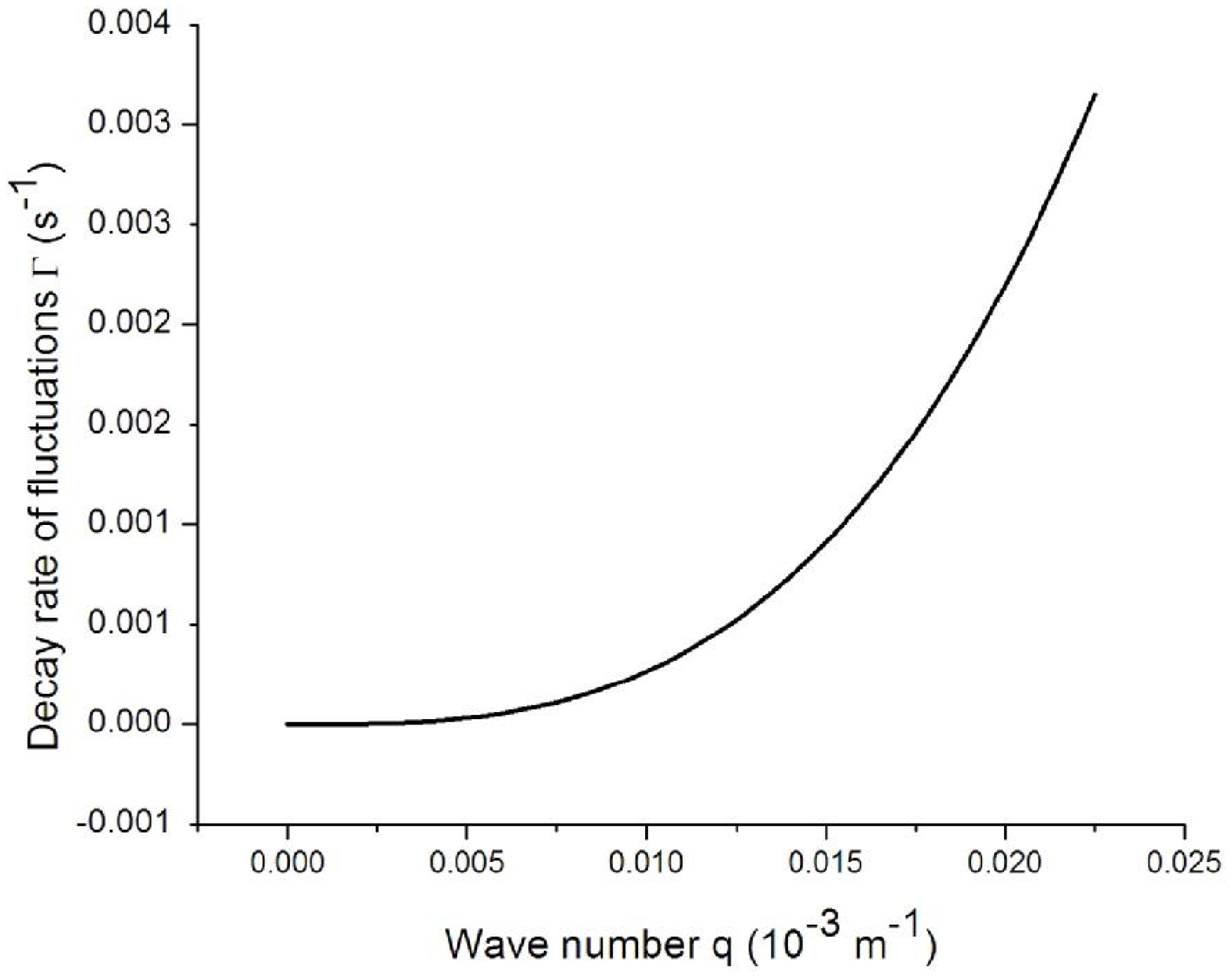,width=18cm,angle=0}}
\caption{Decay rate of fluctuations $\Gamma$ as a function of the wave number. This work predicts a slight deviation from the linear response parabolic behavior. It is noticeable the presence of a plateau in the low wave number region which possibly represents a macroscopic effect of spatial smoothing down of fluctuations.}
\label{Figure 3}
\end{figure}


\begin{thebibliography}{99}
\bibitem{expe} In experimental studies thermostating is usually carried out by means of a so-called proportional controller in order to ensure thermal stability. Thermal decay thus remains proportional for each cycle and so the time-integrated effect becomes exponential.
\end{thebibliography}
\end{document}